\documentclass{article}

\usepackage{amsmath,amsthm,amsfonts,amssymb}

\newtheorem{teo}{Theorem}
\newtheorem{coro}[teo]{Corollary}
\newtheorem{lema}[teo]{Lemma}
\newtheorem{prop}[teo]{Proposition}
\newtheorem{defi}[teo]{Definition}
\newtheorem{obs}[teo]{Remark}

\begin{document}

\title{{On a Selberg-Schur integral}}

\author{S.~M.~Iguri\footnote{e-mail: siguri@iafe.uba.ar}}

\date{\small Instituto de Astronom\'{\i}a y F\'{\i}sica del Espacio (CONICET-UBA).\\ C.~C.~67 - Suc.~28, 1428 Buenos Aires, Argentina. \\ and \\ Dpto. de F\'\i sica, FCEyN, Universidad de Buenos Aires. \\ Ciudad Universitaria Pab. I, 1428 Buenos Aires, Argentina.}

\maketitle

\begin{abstract}
A generalization of Selberg's beta integral involving Schur polynomials associated with partitions with entries not greater than $2$ is explicitly computed. The complex version of this integral is given after proving a general statement concerning the complex extensions of Selberg-Schur integrals. All these results have interesting applications in both mathematics and physics, particularly in conformal field theory, since the conformal blocks for the $SL(2,\mathbb{R})$ Wess-Zumino-Novikov-Witten model can be obtained by analytical continuation of these integrals.
\end{abstract}


{\small \noindent Selberg integral. Schur polynomials. Aomoto integrals.}

\section{Introduction}

Since Selberg's seminal paper \cite{selberg}, Selberg integral and its generalizations have had a deep impact in several areas of both mathematics and physics. In order to visualize their relevance let us mention that Selberg integral was used to prove the Mehta-Dyson conjecture and other cases of the Macdonald conjectures \cite{askey,macdonald,morris} and that some of their extensions have played a central role in the study of some $q$-analogues of constant term identities \cite{evans,kadell1,kadell2,kaneko,stembridge,warnaar}, in the Calogero-Sutherland quantum many body models \cite{awata,forrester1}, in multivariable orthogonal polynomial theory \cite{opdam,stokman}, in random matrix theory \cite{forrester2,keating1,keating2} and in conformal field and string theories \cite{dotsenko,etingof,felder,mimachi,tarasov,tsuchiya}. See \cite{andrews,forrester3} for more details as well as for many other applications and a complete and list of references.

The purpose of this article is to compute some Selberg-type integrals involving Schur polynomials, i.e., integrals of the form $J(\lambda) \equiv J^{(N)}(a,b,\rho;\lambda)$:
\begin{eqnarray}
\label{integral}
J(\lambda) = \int_{\Lambda} \, s_{\lambda}(y_1,\cdots,y_N) \, \prod_{i=1}^N y_i^{a-1} (1-y_i)^{b-1} \prod_{i<j}^N |y_i-y_j|^{2\rho} \, dy_1\wedge\cdots\wedge dy_N,
\end{eqnarray}
where $s_{\lambda}(y_1,\cdots,y_N)$ is the Schur polynomial associated with the partition $\lambda$, the integral is taken over the $N$-dimensional open domain\footnote{\label{footnote}Selberg-type integrals are sometimes defined over the $N$-dimensional simplex $\left\{ (y_1,\dots,y_N) \in \mathbb{R}^N / 0 < y_1 < \cdots < y_N < 1 \right\}$. The symmetry of the integrand under the permutation of any pair of variables implies that these integrals simply differ, if defined over $\Lambda$, in a factor $(1/N!)$.} $\Lambda = (0,1)^N$ and $a$, $b$ and $\rho$ are complex numbers. Throughout the rest of the paper we will follow \cite{aomoto1} denoting by $\Phi(y)$ the function $\Phi(y) \equiv \Phi(a,b,\rho;y_1,\dots,y_N) = \prod_{i=1}^N y_i^{a-1} (1-y_i)^{b-1} \prod_{i<j}^N |y_i-y_j|^{2\rho}$ and by $dy$ the $N$-form $dy_1\wedge\cdots\wedge dy_N$. We will write $s_{\lambda}(y)$ instead of $s_{\lambda}(y_1,\cdots,y_N)$. Furthermore, if no reference is made concerning the integration domains, all real integrals should be understood as over $\Lambda$ and complex integrals over the whole complex plane. 

The case $\lambda=0$ corresponds to Selberg's celebrated formula \cite{selberg}:
\begin{eqnarray}
J(0) = \int \Phi(y) \, dy = \prod_{i=1}^{N} \frac{\Gamma\left(a+(N-i)\rho \right)\Gamma\left(b+(N-i)\rho \right)\Gamma(i\rho+1)}{\Gamma\left(a+b+(2N-i-1)\rho\right)\Gamma(\rho+1)}, \nonumber
\end{eqnarray}
which is absolutely convergent when
\begin{eqnarray}
\label{cond1}
\mbox{Re}(a)>0,\qquad \mbox{Re}(b) > 0 \qquad \mbox{and} \qquad \mbox{Re}(\rho) > - \min \left\{\frac{1}{N},\,\frac{\mbox{Re}(a)}{N-1},\,\frac{\mbox{Re}(b)}{N-1}\right\}.
\end{eqnarray}
This last inequality must be replaced by $\mbox{Re}(\rho) > -1$ when $N=1$. From now on we will assume that these conditions are always fulfilled.

When $\lambda$ is a partition with entries not greater than $1$, namely, $\lambda=(1^{m})$ with $m \le N$, Schur polynomials reduce to elementary symmetric polynomials:
\begin{eqnarray}
s_{(1^m)}(y) \equiv e_{m}(y)=\frac{1}{m!(N-m)!} \, \sum_{\sigma_N}\prod_{i=1}^{m} y_{\sigma_N(i)}, \nonumber
\end{eqnarray}
where the sum runs over all the permutations of the set $\{1,\dots,N\}$ and $e_0(y)=1$. The corresponding Selberg-type integral has been computed by Aomoto in \cite{aomoto1}. It is given by
\begin{eqnarray}
\label{aomoto}
J(1^{m}) = \int e_{m}(y) \, \Phi(y) \, dy = {N \choose m} J(0) \prod_{i=1}^{m} \frac{a+(N-i)\rho}{a+b+(2N-i-1)\rho}.
\end{eqnarray}

A further generalization of Selberg integral has been computed by Kadell \cite{kadell3} within the framework of Jack polynomials theory giving an analytical expression for (\ref{integral}) in the particular case in which $\rho=1$ (no restrictions on $\lambda$). Kadell's integral reads:
\begin{eqnarray}
\int P_{\lambda}^{(1/\rho)}(y) \, \Phi(y) \, dy = P_{\lambda}^{(1/\rho)}(1^N) \, J(0) \, \frac{[a+(N-1)\rho]_{\lambda}^{(\rho)}}{[a+b+2(N-1)\rho]_{\lambda}^{(\rho)}}, \nonumber
\end{eqnarray}
where $P_{\lambda}^{(1/\rho)}(y)$ is a Jack polynomial, $(1^N)$ is a shorthand for $y_1=\cdots=y_N=1$ and $[a]_{\lambda}^{(\rho)}$ is a generalized Pochhammer symbol defined as
\begin{eqnarray}
[a]_{\lambda}^{(\rho)}= \prod_{i\ge 1} (b+(1-i)\rho)_{\lambda_i}, \nonumber
\end{eqnarray}
where the product index runs up to the length of $\lambda$ and $(a)_n$ denotes the usual Pochhammer symbol, i.e., $(a)_n=a(a+1)\cdots(a+n-1)$. When $\rho=1$, Jack polynomials reduce to Schur polynomials, leading us to
\begin{eqnarray}
\label{kadell2}
J(\lambda) = s_{\lambda}(1^N) \, J(0) \, \frac{[a+N-1]_{\lambda}^{(1)}}{[a+b+2N-2]_{\lambda}^{(1)}}.
\end{eqnarray}

In this paper we will compute the integral (\ref{integral}) when $\lambda$ has entries not greater than $2$, i.e., when $\lambda$ is of the form $(2^{n}1^{m-n})$ with $n \le m \le N$, and no restrictions but (\ref{cond1}) are imposed on the other parameters. With this notation $m$ equals the length of $\lambda$ and $|\lambda|=n+m$. We will prove in Theorem 7 that, in this case,
\begin{eqnarray}
&& J(\lambda) = m_{\lambda}(1^N) \, J(0) \, \frac{[a+(N-1)\rho]^{(\rho)}_{\lambda}}{[a+b+2(N-1)\rho]^{(\rho)}_{\lambda}} \, \frac{[a+b+(N-2)\rho]^{(\rho)}_{n}}{[a+b+(2N-m-2)\rho]^{(\rho)}_{n}} \nonumber \\
&& ~~~~~ \times {_4}F_3\left[ \begin{array}{c}
-n, \, -N+m, \, \alpha+\beta+\gamma+2N-n-1, \, \alpha+N-n+1 \\ \alpha+\beta+N-n-1, \, \alpha+\gamma+N-n, \, m-n+2 \end{array} \right], \nonumber
\end{eqnarray}
where $\alpha=a/\rho$, $\beta=b/\rho$ and $\gamma=1/\rho$, $[a]^{(\rho)}_{n}$ is a shorthand for $[a]^{(\rho)}_{(1^n)}$, the hypergeometric series $_4F_3$ is evaluated at $1$ and $m_{\lambda}$ refers to the monomial symmetric polynomial associated to $\lambda$.

We will also compute the complex version of this integral. We will follow similar contour manipulations as those performed by Dotsenko and Fateev in \cite{dotsenko,dotsenko2} for obtaining the following complex generalization of Selberg integral:
\begin{eqnarray}
\label{dotsenko}
\mathcal{J}(0,0) = \int \Phi(z) \Phi(\bar{z}) \, dz \, d\bar z = \frac{1}{N!} \, J(0)^2 \prod_{i=1}^{N} \frac{\mathbf{s}(a+(N-i)\rho) \, \mathbf{s}(b+(N-i)\rho)  \, \mathbf{s}(i\rho)}{\mathbf{s}(a+b+(2N-i-1)\rho) \, \mathbf{s}(\rho)},
\end{eqnarray}
where $dz \, d\bar z = (i/2)^N \, dz_1 \wedge d\bar{z}_1 \cdots dz_N \wedge d\bar{z}_N$ and $\mathbf{s}(z)=\sin(\pi z)$ for all $z \in \mathbb{C}$. This is an absolutely convergent integral provided (\ref{cond1}) is supplemented by
\begin{eqnarray}
\mbox{Re}(a+b+(N-1)\rho)<1 \qquad \mbox{and} \qquad \mbox{Re}(a+b+(2N-2)\rho) < 1. \nonumber
\end{eqnarray}
When dealing with complex integrals we will always assume that these conditions are verified. Let us mention that Aomoto has also obtained Eq.~(\ref{dotsenko}) in \cite{aomoto2} in a formally equivalent way.

Dotsenko and Fateev's techniques have been also used in \cite{iguri} for computing the complex version of Aomoto integral in the context of the computation of winding violating correlation functions for the $SL(2,\mathbb{R})$ Wess-Zumino-Novikov-Witten model in the free field approach. We quote here the result:
\begin{eqnarray}
\label{iguri}
&& \mathcal{J}((1^{m}),(1^{\bar m})) = \int e_{m}(z) \, e_{\bar m}(\bar{z}) \, \Phi(z) \Phi(\bar{z}) \, dz \, d\bar z \nonumber \\
&& ~~~~~ = \frac{1}{N!} \, J(1^{m}) J(1^{\bar m}) \prod_{i=1}^{N} \frac{\mathbf{s}(a+(N-i)\rho) \, \mathbf{s}(b+(N-i)\rho) \,  \mathbf{s}(i\rho)}{\mathbf{s}(a+b+(2N-i-1)\rho) \, \mathbf{s}(\rho)} \nonumber \\
&& ~~~~~ = {N \choose m} {N \choose \bar m} \mathcal{J}(0,0) \prod_{i=1}^{m} \frac{a+(N-i)\rho}{a+b+(2N-i-1)\rho} \prod_{i=1}^{\bar m} \frac{a+(N-i)\rho}{a+b+(2N-i-1)\rho}. \nonumber
\end{eqnarray}

The complex version of the integral we will compute in this article is closely related to the conformal blocks of the same theory. Actually, within the framework of the Coulomb gas method, and after computing the $\beta$-$\gamma$ contribution, the factorization limit winding conserving four-point function for the $SL(2,\mathbb{R})$ Wess-Zumino-Novikov-Witten model can be entirely expressed in terms of these integrals (see \cite{nunez2} for more details). As a corollary of Theorem 14 we will obtain
\begin{eqnarray}
&& \mathcal{J}(\lambda,\bar{\lambda}) = \int s_{\lambda}(z) \, s_{\bar{\lambda}}(\bar{z}) \, \Phi(z) \Phi(\bar{z}) \, dz \, d\bar z \nonumber \\ 
&& ~~~~~ = \frac{1}{N!} \, m_{\lambda}(1^N) \, m_{\bar{\lambda}}(1^N) \, \mathcal{J}(0,0) \, \frac{[a+(N-1)\rho]^{(\rho)}_{\lambda}}{[a+b+2(N-1)\rho]^{(\rho)}_{\lambda}} \,  \frac{[a+(N-1)\rho]^{(\rho)}_{\bar{\lambda}}}{[a+b+2(N-1)\rho]^{(\rho)}_{\bar{\lambda}}} \nonumber \\
&& ~~~~~ \times \frac{[a+b+(N-2)\rho]^{(\rho)}_{n}}{[a+b+(2N-m-2)\rho]^{(\rho)}_{n}} \, \frac{[a+b+(N-2)\rho]^{(\rho)}_{\bar{n}}}{[a+b+(2N-\bar m-2)\rho]^{(\rho)}_{\bar n}}  \nonumber \\
&& ~~~~~ \times {_4}F_3\left[ \begin{array}{c}
-n, \, -N+m, \, \alpha+\beta+\gamma+2N-n-1, \, \alpha+N-n+1 \\ \alpha+\beta+N-n-1, \, \alpha+\gamma+N-n, \, m-n+2 \end{array} \right] \nonumber \\
&& ~~~~~ \times {_4}F_3\left[ \begin{array}{c}
-\bar n, \, -N+\bar m, \, \alpha+\beta+\gamma+2N-\bar n-1, \, \alpha+N-\bar n+1 \\ \alpha+\beta+N-\bar n-1, \, \alpha+\gamma+N-\bar n, \, \bar m-\bar n+2 \end{array} \right], \nonumber
\end{eqnarray}
where $\lambda=(2^n1^{m-n})$ and $\bar{\lambda}=(2^{\bar n}1^{\bar m - \bar n})$.

The plan of the paper is as follows. In the next section we will present several lemmas that will allow us to find in section \ref{sec3} a double-sided recurrence relation from which the explicitly analytic expression for the integral (\ref{integral}) with $\lambda=(2^{n}1^{m-n})$ follows. In section \ref{sec4} we derive the complex version of this result.

\section{Preliminary lemmas}
\label{sec2}

In this section we are going to derive some formulas that will be useful for proving our main theorems.

\begin{lema}
Let $m$, $n$ and $k$ be three integer numbers such that $0 \le n < m \le N$ and $1 < k \le N$. Then,
\begin{eqnarray}
\label{lemma1}
 \int \frac{y_1^2 \cdots y_{n+1}^2y_{n+2} \cdots y_{m}}{y_1 - y_k} \, \Phi(y) \, dy = \left\{ \begin{array}{lll} 0 & \mbox{if} & 2 \le k \le n+1 \\\frac{1}{2} \int y_1^2 \cdots y_{n}^2y_{n+1} \cdots y_{m} \, \Phi(y) \, dy & \mbox{if} & n+2 \le k \le m \\ \int y_1^2 \cdots y_{n}^2y_{n+1} \cdots y_{m} \, \Phi(y) \, dy & \mbox{if} & m+1 \le k \le N. \end{array} \right.
\end{eqnarray}
\end{lema}

\noindent\textit{Proof.} When $2 \le k \le n+1$ the sign of the integrand changes if $y_1$ and $y_k$ are interchanged. Since $\Lambda$ is invariant under the permutation of any pair of indices $1,\dots,N$ and $\Phi(y)$ is a symmetric function of $y_1,\dots,y_N$, this proves the first equality in (\ref{lemma1}).

When $n+2 \le k \le m$ and under the same transposition we get
\begin{eqnarray}
\label{lemma1.case2}
\frac{y_1^2y_k}{y_1-y_k} \leftrightarrow \frac{y_1y_k^2}{y_k-y_1} = y_1 y_k - \frac{y_1^2y_k}{y_1-y_k},
\end{eqnarray}
and the left-hand side of (\ref{lemma1}) gives
\begin{eqnarray}
\int y_1^2 \cdots y_{n}^2y_{n+1} \cdots y_{m} \, \Phi(y) \, dy - \int \frac{y_1^2 \cdots y_{n+1}^2y_{n+2} \cdots y_{m}}{y_k - y_1} \, \Phi(y) \, dy. \nonumber
\end{eqnarray}
This shows that the second equality in (\ref{lemma1}) holds.

Finally, if $m+1 \le k \le N$, the transposition $y_1 \leftrightarrow y_k$ gives
\begin{eqnarray}
\frac{y_1^2}{y_1-y_k} \leftrightarrow \frac{y_k^2}{y_k-y_1} = y_1 + y_k - \frac{y_1^2}{y_1-y_k}. \nonumber
\end{eqnarray}
It follows that the left-hand side of (\ref{lemma1}) equals
\begin{eqnarray}
2 \int y_1^2 \cdots y_{n}^2y_{n+1} \cdots y_{m} \, \Phi(y) \, dy - \int \frac{y_1^2 \cdots y_{n+1}^2y_{n+2} \cdots y_{m}}{y_k - y_1} \, \Phi(y) \, dy, \nonumber
\end{eqnarray}
completing the proof of the lemma. \qed

\begin{lema}
Let $m$ and $n$ be two integer numbers such that $0 \le n < m \le N$. Then,
\begin{eqnarray}
\label{lemma2}
&& (b-1) \int \frac{y_2^2 \cdots y_{n+1}^2y_{n+2} \cdots y_{m}}{1-y_1} \, \Phi(y) \, dy =(b-1) \int y_1^2 \cdots y_{n}^2y_{n+1} \cdots y_{m-1} \, \Phi(y) \, dy\nonumber \\
&& ~~~~~ +\left[a+b+(2N-n-m-1)\rho \right] \int y_1^2 \cdots y_{n}^2y_{n+1} \cdots y_{m} \, \Phi(y) \, dy.
\end{eqnarray}
\end{lema}

\noindent\textit{Proof.} Since $\Phi(y)$ vanishes at the boundary values $y_1=0$ and $y_1=1$, it follows from Stokes' theorem that
\begin{eqnarray}
\label{lemma2.stokes}
&& 0 = \int d \left( y_1^2 \cdots y_{n+1}^2y_{n+2} \cdots y_{m} \, \Phi(y) \, dy' \right) = 2 \rho \sum_{k=2}^N \int \frac{y_1^2 \cdots y_{n+1}^2y_{n+2} \cdots y_{m}}{y_1 - y_k} \, \Phi(y) \, dy \nonumber \\
&& ~~~~~ + (a+1) \int y_1^2 \cdots y_{n}^2y_{n+1} \cdots y_{m} \, \Phi(y) \, dy - (b-1) \int \frac{y_1^2 \cdots y_{n+1}^2y_{n+2} \cdots y_{m}}{1-y_1} \, \Phi(y) \, dy,
\end{eqnarray}
where $dy'$ denotes the $(N-1)$-form $dy_2 \wedge \cdots \wedge dy_N$.

Using
\begin{eqnarray}
\frac{y_1^2}{1-y_1} = \frac{1}{1-y_1}-y_1-1, \nonumber
\end{eqnarray}
it is straightforward to check that
\begin{eqnarray}
\label{lemma2bis}
&& \int \frac{y_1^2 \cdots y_{n+1}^2y_{n+2} \cdots y_{m}}{1-y_1} \, \Phi(y) \, dy = \int \frac{y_2^2 \cdots y_{n+1}^2y_{n+2} \cdots y_{m}}{1-y_1} \, \Phi(y) \, dy \nonumber \\
&& ~~~~~ - \int y_1^2 \cdots y_{n}^2y_{n+1} \cdots y_{m} \, \Phi(y) \, dy - \int y_1^2 \cdots y_{n}^2y_{n+1} \cdots y_{m-1} \, \Phi(y) \, dy.
\end{eqnarray}

On the other hand, we have from (\ref{lemma1}),
\begin{eqnarray}
\label{coro2}
\sum_{k=2}^{N} \int \frac{y_1^2 \cdots y_{n+1}^2y_{n+2} \cdots y_{m}}{y_1 - y_k} \, \Phi(y) \, dy = \frac{1}{2} \, (2N-n-m-1)\int y_1^2 \cdots y_{n}^2y_{n+1} \cdots y_{m} \, \Phi(y) \, dy.
\end{eqnarray}

Eq.~(\ref{lemma2}) follows after replacing (\ref{lemma2bis}) and (\ref{coro2}) in (\ref{lemma2.stokes}). \qed

\begin{lema}
Let $m$, $n$ and $k$ be integers verifying $0 \le n < m \le N$ and $2 \le k \le N$. Then,
\begin{eqnarray}
\label{lemma3}
\int \frac{y_1y_2^2 \cdots y_{n+1}^2y_{n+2} \cdots y_{m}}{y_1 - y_k} \, \Phi(y) \, dy = \left\{ \begin{array}{lll}
-\frac{1}{2} \int y_1^2 \cdots y_{n-1}^2y_{n} \cdots y_{m} \, \Phi(y) \, dy & \mbox{if} & 2 \le k \le n+1 \\ 0 & \mbox{if} & n+2 \le k \le m \\ \frac{1}{2} \int y_1^2 \cdots y_{n}^2y_{n+1} \cdots y_{m-1} \, \Phi(y) \, dy & \mbox{if} & m+1 \le k \le N. \end{array}\right.
\end{eqnarray}
\end{lema}

\noindent\textit{Proof.} Consider the transposition $y_1 \leftrightarrow y_k$. When $2 \le k \le n+1$ it follows from (\ref{lemma1.case2}) that the left-hand side of (\ref{lemma3}) equals
\begin{eqnarray}
- \int y_1^2 \cdots y_{n-1}^2y_{n} \cdots y_{m} \, \Phi(y) \, dy - \int \frac{y_1y_2^2 \cdots y_{n+1}^2y_{n+2} \cdots y_{m}}{y_1 - y_k} \, \Phi(y) \, dy, \nonumber
\end{eqnarray}
showing that the first equality in (\ref{lemma3}) holds.

If $n+2 \le k \le m$ the same transposition changes the sign of the integrand proving thus that (\ref{lemma3}) vanishes.

Finally, if $m+1 \le k \le N$, the transposition $y_1 \leftrightarrow y_k$ gives
\begin{eqnarray}
\frac{y_1}{y_1 - y_k} \leftrightarrow \frac{y_k}{y_k - y_1} = 1-\frac{y_1}{y_1 - y_k}, \nonumber
\end{eqnarray}
and the left-hand side of (\ref{lemma3}) equals
\begin{eqnarray}
\int y_1^2 \cdots y_{n}^2y_{n+1} \cdots y_{m-1} \, \Phi(y) \, dy - \int \frac{y_1y_2^2 \cdots y_{n+1}^2y_{n+2} \cdots y_{m}}{y_1 - y_k} \, \Phi(y) \, dy. \nonumber
\end{eqnarray}
This completes the proof of the lemma. \qed

\begin{lema}
Let $m$ and $n$ be two integers such that $0 \le n < m \le N$. Then,
\begin{eqnarray}
\label{lemma4}
&& \left[a+b+(2N-n-m-1)\rho \right] \int y_1^2 \cdots y_{n}^2y_{n+1} \cdots y_{m} \, \Phi(y) \, dy = \nonumber \\
&& ~~~~~ = \left[a+(N-m)\rho \right] \int y_1^2 \cdots y_{n}^2y_{n+1} \cdots y_{m-1} \, \Phi(y) \, dy - n\rho \int y_1^2 \cdots y_{n-1}^2y_{n} \cdots y_{m} \, \Phi(y) \, dy.
\end{eqnarray}
\end{lema}

\noindent\textit{Proof.} Applying again Stokes' theorem we obtain
\begin{eqnarray}
\label{lemma4.stokes}
&& 0 = \int d\left(y_1 y_2^2 \cdots y_{n+1}^2y_{n+2} \cdots y_{m} \, \Phi(y) \, dy' \right) = 2 \rho \sum_{k=2}^N \int \frac{y_1y_2^2 \cdots y_{n+1}^2y_{n+2} \cdots y_{m}}{y_1 - y_k} \, \Phi(y) \, dy \nonumber \\
&& ~~~~~+ (a+b-1) \int y_1^2 \cdots y_{n}^2y_{n+1} \cdots y_{m-1} \, \Phi(y) \, dy - (b-1) \int \frac{y_2^2 \cdots y_{n+1}^2y_{n+2} \cdots y_{m}}{1-y_1} \, \Phi(y) \, dy.
\end{eqnarray}

On the other hand, from (\ref{lemma3}) we get
\begin{eqnarray}
\label{coro2b}
&& \sum_{k=2}^{N} \int\frac{y_1y_2^2 \cdots y_{n+1}^2y_{n+2} \cdots y_{m}}{y_1 - y_k} \, \Phi(y) \, dy = - \frac{1}{2} \, n \int y_1^2 \cdots y_{n-1}^2y_{n} \cdots y_{m} \, \Phi(y) \, dy \nonumber \\
&& ~~~~~ + \frac{1}{2} \, (N-m) \int y_1^2 \cdots y_{n}^2y_{n+1} \cdots y_{m-1} \, \Phi(y) \, dy.
\end{eqnarray}

The proof of the lemma is straightforward after replacing (\ref{lemma2}) and (\ref{coro2b}) in (\ref{lemma4.stokes}). \qed

\section{Main theorems}
\label{sec3}

The equation in Lemma 4 defines a double recursive relation generalizing a formula previously obtained by Aomoto in \cite{aomoto1} when computing Selberg-Schur integrals associated with elementary partitions. Its solution will allow us to find an analytical expression for the integral (\ref{integral}) when $\lambda=(2^n1^{m-n})$.

\begin{teo} Let $m$ and $n$ be two integers such that $0 \le n \le m \le N$ and let $\lambda=(2^n1^{m-n})$. The following identity holds:
\begin{eqnarray}
\label{casi}
&& \int y_1^2 \cdots y_{n}^2y_{n+1} \cdots y_{m} \, \Phi(y) \, dy = J(0) \frac{[a+(N-1)\rho]^{(\rho)}_{\lambda}}{[a+b+2(N-1)\rho]^{(\rho)}_{\lambda}} \, \frac{[a+b+(N-2)\rho]^{(\rho)}_{n}}{[a+b+(2N-m-2)\rho]^{(\rho)}_{n}} \nonumber \\
&& ~~~~~ \times {_3}F_2\left[ \begin{array}{c}
-n, \, -N+m, \, \alpha+\beta+\gamma+2N-n-1 \\ \alpha+\beta+N-n-1, \, \alpha+\gamma+N-n\end{array} \right].
\end{eqnarray}
\end{teo}

\noindent\textit{Proof.} First let us define:
\begin{eqnarray}
\label{filn}
\Psi(m,n) = \frac{1}{\rho^n n! J(0)} \, \frac{\prod_{i=1}^{m+n} [a+b+(2N-i-1)\rho]}{ \prod_{i=1}^{m} [a+(N-i)\rho]} \int y_1^2 \cdots y_{n}^2y_{n+1} \cdots y_{m} \, \Phi(y) \, dy.
\end{eqnarray}
Multiplying both sides of (\ref{lemma4}) by $\prod_{i=1}^{m+n-1} [a+b+(2N-i-1)\rho] /\rho^n n! J(0) \prod_{i=1}^{m} [a+(N-i)\rho]$ we get the following recursion for $\Psi(m,n)$:
\begin{eqnarray}
\label{recurrence2}
\Psi(m,n) = \Psi(m-1,n) - \Psi(m,n-1),
\end{eqnarray}
for $0 \le n < m \le N$, where, by definition, $\Psi(m,-1)=0$ for $m=1,2,\dots,N$.

The boundary values of this recurrence can be obtained using Aomoto integral. In fact, it is straightforward to see from (\ref{aomoto}) that $\Psi(m,0)=1$ for all $m \ge 0$. On the other hand, $\Psi(N,n)$ for $n=0,1,\dots,N$ can also be computed in terms of Aomoto integral. We have
\begin{eqnarray}
 J'(1^n) = {N \choose n} \int y_1 \cdots y_{n} \, \Phi'(y) \, dy = {N \choose n} \int y_1^2 \cdots y_{n}^2y_{n+1} \cdots y_{N} \, \Phi(y) \, dy, \nonumber
\end{eqnarray}
where we have introduced a dash whenever the value of the parameter $a$ is increased in a unit, namely, when $a \mapsto a'=a+1$. Then, using $J'(0)=J(1^N)$ it follows that
\begin{eqnarray}
\label{filn22}
\Psi(N,n)= \frac{1}{n! \rho^n} \, \prod_{i=1}^n \frac{[a+b+(N-i-1)\rho][a+1+(N-i)\rho]}{a+b+1+(2N-i-1)\rho}.
\end{eqnarray}

It can be proved inductively that the solution of (\ref{recurrence2}) with these boundary values is given by
\begin{eqnarray}
\Psi(m,n)= \sum_{k = 0}^n {N-m \choose k} \Psi(N,n-k), \nonumber
\end{eqnarray}
which becomes, using (\ref{filn22}),
\begin{eqnarray}
\label{sol}
&& \Psi(m,n)= \sum_{k = 0}^n {N-m \choose k} \prod_{i=1}^{n-k} \frac{[a+b+(N-i-1)\rho][a+1+(N-i)\rho]}{i\rho \,[a+b+1+(2N-i-1)\rho]} \nonumber \\
&& ~~~~~ = \Psi(N,n) \sum_{k = 0}^n {N-m \choose k} \prod_{i=1}^{k} \frac{(n-i+1)\rho [a+b+1+(2N-n+i-2)\rho]}{[a+b+(N-n+i-2)\rho][a+1+(N-n+i-1)\rho]}.
\end{eqnarray}

Replacing (\ref{sol}) in (\ref{filn}) we get
\begin{eqnarray}
\label{linda}
&& \int y_1^2 \cdots y_{n}^2y_{n+1} \cdots y_{m} \, \Phi(y) \, dy = J(0) \frac{[a+(N-1)\rho]^{(\rho)}_{\lambda}}{[a+b+2(N-1)\rho]^{(\rho)}_{\lambda}} \nonumber \\
&& ~~~~~ \times \prod_{i=1}^n \frac{a+b+(N-i-1)\rho}{a+b+(2N-m-i-1)\rho} \, {_3}F_2\left[ \begin{array}{c}
-n, \, -N+m, \, \alpha+\beta+\gamma+2N-n-1 \\ \alpha+\beta+N-n-1, \, \alpha+\gamma+N-n\end{array} \right].
\end{eqnarray}

Expression (\ref{casi}) is finally obtained when the product in (\ref{linda}) is rewritten as a quotient of Pochhammer symbols. \qed

\begin{obs}
Notice that (\ref{casi}) can be entirely expressed in terms of hypergeometric series. Using Chu-Vandermonde's theorem it is straightforward to check that
\begin{eqnarray}
\prod_{i=1}^n \frac{a+b+(2N-m-i-1)\rho}{a+b+(N-i-1)\rho} = \frac{(\alpha+\beta+2N-n-m-1)_n}{(\alpha+\beta+N-n-1)_n} = {_2}F_1\left[ \begin{array}{c}
-n, \, -N+m \\ \alpha+\beta+N-n-1\end{array} \right], \nonumber
\end{eqnarray}
and then,
\begin{eqnarray}
&& \int y_1^2 \cdots y_{n}^2y_{n+1} \cdots y_{m} \, \Phi(y) \, dy \nonumber \\
&& ~~~~~ = J(0) \frac{[a+(N-1)\rho]^{(\rho)}_{\lambda}}{[a+b+2(N-1)\rho]^{(\rho)}_{\lambda}} \, \frac{{_3}F_2\left[ \begin{array}{c}
-n, \, -N+m, \, \alpha+\beta+\gamma+2N-n-1 \\ \alpha+\beta+N-n-1, \, \alpha+\gamma+N-n\end{array} \right]}{{_2}F_1\left[ \begin{array}{c}
-n, \, -N+m \\ \alpha+\beta+N-n-1\end{array} \right]}. \nonumber
\end{eqnarray}
\end{obs}

\begin{teo} Let $\lambda$ be any partition with entries not greater than $2$, namely, $\lambda=(2^n1^{m-n})$. Then,
\begin{eqnarray}
\label{incre}
&& J(\lambda) = m_{\lambda}(1^N) \, J(0) \, \frac{[a+(N-1)\rho]^{(\rho)}_{\lambda}}{[a+b+2(N-1)\rho]^{(\rho)}_{\lambda}} \, \frac{[a+b+(N-2)\rho]^{(\rho)}_{n}}{[a+b+(2N-m-2)\rho]^{(\rho)}_{n}} \nonumber \\
&& ~~~~~ \times {_4}F_3\left[ \begin{array}{c}
-n, \, -N+m, \, \alpha+\beta+\gamma+2N-n-1, \, \alpha+N-n+1 \\ \alpha+\beta+N-n-1, \, \alpha+\gamma+N-n, \, m-n+2 \end{array} \right].
\end{eqnarray}
\end{teo}

\noindent\textit{Proof.} First recall that any Schur polynomial can be expanded in term of monomial symmetric polynomials as
\begin{eqnarray}
s_{\lambda}(y) = \sum_{\mu \preceq \lambda} K_{\lambda\mu} m_{\mu}(y), \nonumber
\end{eqnarray}
where $K_{\lambda\mu}$ is the Kostka number associated with $\lambda$ and $\mu$ and $m_{\mu}(y)$ is the monomial symmetric polynomial indexed by $\mu$. The order in the sum is the usual dominance ordering on partitions.

In our case, $\lambda=(2^n1^{m-n})$, then $\mu \preceq \lambda$ if and only if $\mu=(2^{n-r}1^{m-n+2r})$ for $r=0,1,\dots,n$.

On the other hand we have $K_{\lambda\mu} = K_{(2^{r}1^{m-n})(1^{m-n+2r})}$ which is the number of standard Young tableaux of shape $(2^{r}1^{m-n})$ (cf. \cite{stanley}). This number is given by
\begin{eqnarray}
K_{\lambda\mu} = {m-n+2r \choose m-n+r} - {m-n+2r \choose m-n+r+1} = \frac{m-n+1}{m-n+2r+1} \, {m-n+2r+1 \choose r}. \nonumber
\end{eqnarray}

It follows that
\begin{eqnarray}
\label{schur2cc}
&& J(\lambda)= \sum_{r=0}^n \frac{m-n+1}{m-n+2r+1} \, {m-n+2r+1 \choose r} \int m_{\mu}(y) \, \Phi(y) \, dy .
\end{eqnarray}

Since
\begin{eqnarray}
\int m_{\mu}(y) \, \Phi(y) \, dy = m_{\mu}(1^N) \, \int y_1^2 \cdots y_{n-r}^2 y_{n-r+1} \cdots y_{m+r} \, \Phi(y) \, dy, \nonumber
\end{eqnarray}
where $m_{\mu}(1^N)=N!/(n-r)!(m-n+2r)!(N-m-r)!$, after replacing (\ref{linda}) in (\ref{schur2cc}) we get
\begin{eqnarray}
&& J(\lambda) = J(0) \sum_{r=0}^n \frac{(m-n+1)N!}{r!(m-n+r+1)!(n-r)!(N-m-r)!} \nonumber \\
&& ~~~~~ \times \frac{[a+(N-1)\rho]^{(\rho)}_{\mu}}{[a+b+2(N-1)\rho]^{(\rho)}_{\mu}} \prod_{i=1}^{n-r} \frac{a+b+(N-i-1)\rho}{a+b+(2N-m-r-i-1)\rho} \nonumber \\
&& ~~~~~ \times {_3}F_2\left[ \begin{array}{c}
-n+r, \, -N+m+r, \, \alpha+\beta+\gamma+2N-n-1+r \\ \alpha+\beta+N-n-1+r, \, \alpha+\gamma+N-n+r\end{array} \right]. \nonumber
\end{eqnarray}
%

Explicitly writing the hypergeometric series we obtain
\begin{eqnarray}
&& J(\lambda) = m_{\lambda}(1^N) \, J(0) \, \frac{[a+(N-1)\rho]^{(\rho)}_{\lambda}}{[a+b+2(N-1)\rho]^{(\rho)}_{\lambda}} \, \, \frac{[a+b+(N-2)\rho]^{(\rho)}_{n}}{[a+b+(2N-m-2)\rho]^{(\rho)}_{n}} \nonumber \\
&& ~~~~~ \times \sum_{r=0}^n \sum_{k=0}^{n-r} \frac{(-1)^r}{r!k!}\frac{(1-\alpha-N+m)_r}{(m-n+2)_r} \frac{(-n)_{r+k}(-N+m)_{r+k}(\alpha+\beta+\gamma+2N-n-1)_{r+k}}{(\alpha+\beta+N-n-1)_{r+k}(\alpha+\gamma+N-n)_{r+k}}, \nonumber
\end{eqnarray}
which after a suitable change in the summation indices becomes
\begin{eqnarray}
\label{cuenta5}
&& J(\lambda) = m_{\lambda}(1^N) \, J(0) \, \frac{[a+(N-1)\rho]^{(\rho)}_{\lambda}}{[a+b+2(N-1)\rho]^{(\rho)}_{\lambda}} \, \, \frac{[a+b+(N-2)\rho]^{(\rho)}_{n}}{[a+b+(2N-m-2)\rho]^{(\rho)}_{n}} \nonumber \\
&& ~~~~~ \times \sum_{k=0}^n \sum_{r=0}^{k} \frac{(-k)_r(1-\alpha-N+m)_r}{r!(m-n+2)_r} \frac{(-n)_{k}(-N+m)_{k}(\alpha+\beta+\gamma+2N-n-1)_{k}}{k!(\alpha+\beta+N-n-1)_{k}(\alpha+\gamma+N-n)_{k}}.
\end{eqnarray}

The sum over the index $r$ gives
\begin{eqnarray}
\label{increible}
&& \sum_{r=0}^{k} \frac{(-k)_r(1-\alpha-N+m)_r}{r!(m-n+2)_r} = {_2}F_1\left[ \begin{array}{c}
-k, \, 1-\alpha-N+m \\ m-n+2 \end{array} \right] = \frac{(\alpha+N-n+1)_k}{(m-n+2)_k},
\end{eqnarray}
where we have used again Chu-Vandermonde's theorem.

Eq.~(\ref{incre}) is obtained after replacing (\ref{increible}) in (\ref{cuenta5}). \qed

\begin{obs}
It is interesting to note at this point that simplifications occur when $n=0$ or $m=N$. In both cases the hypergeometric series equal $1$, reducing Eq.~(\ref{incre}) to an Aomoto integral, as expected.
\end{obs}

\begin{obs}
Another simplification occurs when $\rho=1$. In this case the ${_4}F_3$ hypergeometric series in (\ref{incre}) reduces to a balanced ${_3}F_2$ hypergeometric series that can be computed using Pfaff-Saalsch\"utz' theorem, namely,
\begin{eqnarray}
&& {_3}F_2\left[ \begin{array}{c}
-n, \, -N+m, \, a+b+2N-n \\ a+b+N-n-1, \, m-n+2 \end{array} \right] \nonumber \\
&& ~~~~~ = \frac{(N-n+2)_n}{(m-n+2)_n} \, \frac{\Gamma(a+b+N-n-1)\Gamma(a+b+2N-m-1)}{\Gamma(a+b+N-1)\Gamma(a+b+2N-m-n-1)} \nonumber \\
&& ~~~~~ = \frac{(N-n+2)_n}{(m-n+2)_n} \, \frac{[a+b+(2N-m-2)\rho]^{(\rho)}_{n}}{[a+b+(N-2)\rho]^{(\rho)}_{n}}. \nonumber
\end{eqnarray}

When replaced in (\ref{incre}) it gives
\begin{eqnarray}
\label{incre2}
&& J(\lambda) = m_{\lambda}(1^N) \, \frac{(N-n+2)_n}{(m-n+2)_n} \, J(0) \frac{[a+N-1]^{(1)}_{\lambda}}{[a+b+2N-2]^{(1)}_{\lambda}}.
\end{eqnarray}

Let us denote by $u$ any box in the shape of the partition $\lambda$, by $c(u)$ the content of $\lambda$ at $u$ and by $h(u)$ the hook length of $\lambda$ at $u$. Since $s_{\lambda}(1^N) = \prod_{u\in \lambda} (N+c(u))/h(u)$ (cf. \cite{stanley}), we have in our case
\begin{eqnarray}
s_{\lambda}(1^N) = \frac{1}{(m-n)!} \prod_{i=1}^{n} \frac{(N-i+1)(N-i+2)}{i(m-i+2)}\prod_{i=1}^{m-n} (N-n-i+1) = m_{\lambda}(1^N) \, \frac{(N-n+2)_n}{(m-n+2)_n}. \nonumber
\end{eqnarray}
This shows that (\ref{incre2}) reproduces Kadell's Selberg-Jack integral (\ref{kadell2}).
\end{obs}

\section{Complex version}
\label{sec4}

In order to compute the complex version of the integral (\ref{integral}) we will need the following.

\begin{prop}
\label{prop9}
Let $\lambda$ be a partition and let
\begin{eqnarray}
\label{iguri7}
&& J_{q}(\lambda)=\int_0^1 dy_1 \cdots \int_0^1 dy_q \int_{1}^{+\infty} dy_{q+1} \cdots \int_1^{+\infty} dy_N \, s_{\lambda}(y) \, \Phi(y), \nonumber
\end{eqnarray}
for $q=0,1,\dots,N$. Then, the following equality holds:
\begin{eqnarray}
\label{iguri8}
&& J_q(\lambda) = \frac{q}{N-q+1} \, J_{q-1}(\lambda) \, \frac{\mbox{\bf s}(a+b+(N+q-2)\rho)}{\mbox{\bf s}(a+(q-1)\rho)} \, \frac{\mbox{\bf s}((N-q+1)\rho)}{\mbox{\bf s}(q\rho)}.
\end{eqnarray}
\end{prop}

\noindent\textit{Proof.} Consider the following complex integrals:
\begin{eqnarray}
\label{int2}
&& J^{\pm}(\lambda)=\oint_{\mathcal{C}^{\pm}} s_{\lambda}(y;z) \, \Phi(y;z) \, dz, \nonumber
\end{eqnarray}
where $s_{\lambda}(y;z)$ is a shorthand for $s_{\lambda}(y_1,\dots,y_{q-1},z,y_{q+1},\dots,y_N)$, $\Phi(y;z)$ is given by
\begin{eqnarray}
\label{int2bb}
&& \Phi(y;z) = z^{a-1} (1-z)^{b-1} \prod_{{i=1} \atop {i \ne q}}^N (y_i - z)^{2\rho} |y_i|^{a-1} |1-y_i|^{b-1} \prod_{{i<j} \atop {i,j\ne q}}^N |y_i-y_j|^{2\rho}, \nonumber
\end{eqnarray}
and the integrations are made over the paths along the real axis on the upper (resp., lower) complex half plane with semi-circles of vanishing radii around the points $y_1 < \dots < y_{q-1} < y_{q+1} < \dots < y_{N}$ as indentations. We have denoted these contours by $\mathcal{C}^{\pm}$, respectively.

If multivalued power functions are defined in terms of their principal arguments we have that $(y_i-z)^{2\rho}$ equals $|y_i-z|^{2\rho}$ picking up an additional phase factor of the form $\exp{(\mp 2\pi i \rho)}$ whenever $y_i < z$ and $\mbox{Im}(z)>0$ (resp., $\mbox{Im}(z)<0$). With this definition, $\Phi(y;z)$ is an analytic function in the upper (resp., lower) half plane, and since it decays sufficiently fast as $|z| \rightarrow +\infty$ when conditions (\ref{cond1}) are fulfilled, we can freely close the contour and conclude, by means of Cauchy's theorem, that both $J^{+}(\lambda)$ and $J^{-}(\lambda)$ vanish.

On the other hand, in the limit in which the radii of the semi-circles around $y_1,\dots,y_{q-1},y_{q+1},\dots,y_{N}$ shrink to zero, integrals $J^{\pm}(\lambda)$ are given by
\begin{eqnarray}
\label{int3}
&& J^{\pm}(\lambda)=0=e^{\pm i\pi a} \int_{-\infty}^0 s_{\lambda}(y;z) \, \Phi(y;z) \, dz + \sum_{k=0}^{q-1} e^{\mp 2\pi i k \rho} \int_{y'_k}^{y'_{k+1}} s_{\lambda}(y;z) \, \Phi(y;z) \, dz \nonumber \\
&& ~~~~~  - e^{\mp i\pi b} \sum_{k=q}^{N} e^{\mp 2\pi i (k-1) \rho} \int_{y'_k}^{y'_{k+1}} s_{\lambda}(y;z) \, \Phi(y;z) \, dz, \nonumber
\end{eqnarray}
where all the integrals are now over the real line, $y'_0 = 0$, $y'_q = 1$, $y'_{N+1} = +\infty$ and $y'_k=y_k$ for all other values of $k$. It follows that
\begin{eqnarray}
\label{int4}
&& \sum_{k=0}^{q-1} \mbox{\bf s}(a+2 k \rho) \int_{y'_k}^{y'_{k+1}} s_{\lambda}(y) \, \Phi(y) \, dy_q = \sum_{k=q}^{N} \mbox{\bf s}(a+b+ 2(k-1) \rho) \int_{y'_k}^{y'_{k+1}} s_{\lambda}(y) \, \Phi(y) \, dy_q. \nonumber
\end{eqnarray}

Integrating both terms on $\{ (y_1,\dots,y_{q-1}) \in \mathbb{R}^{q-1} / 0 < y_1 < \cdots < y_{q-1} < 1 \}$ and $ \{(y_{q+1},\dots,y_N) \in \mathbb{R}^{N-q} / 1 < y_{q+1} < \cdots < y_N < +\infty \}$ we get
\begin{eqnarray}
\label{int4bis}
&& \frac{1}{p} \, J_{p}(\lambda) \, \sum_{k=0}^{q-1} \mbox{\bf s}(a+2 k \rho) = \frac{1}{N-q+1} \, J_{q-1}(\lambda) \, \sum_{k=q}^{N} \mbox{\bf s}(a+b+ 2(k-1) \rho). \nonumber
\end{eqnarray}

After using the following trigonometric identity:
\begin{eqnarray}
&& \sum_{k=0}^{q-1} \sin(a+2k\rho) = \frac{\sin( q \rho )}{\sin(\rho)} \, \sin(a+(q-1) \rho). \nonumber
\end{eqnarray}
Eq.~(\ref{iguri8}) is finally obtained. \qed

\begin{defi}
Let $\lambda$ be any partition and let $N$ be any positive integer greater than its length. We will define the {\it N-conjugate of $\lambda$} as the partition $\lambda^{N}$ whose shape is the $180^{\circ}$-rotated skew shape $((\lambda_1^N)+(\lambda_N^N))/\lambda$.
\end{defi}

\begin{lema}
\label{lemma5}
Let $\lambda$ be any partition and let $s_{\lambda}(y)$ be the Schur polynomial associated with it. Then,
\begin{eqnarray}
\label{schur1/y}
s_{\lambda}(1/y) = s_{\lambda^N}(y) \, \prod_{i=1}^N y_i^{-\lambda_1-\lambda_N}.
\end{eqnarray}
\end{lema}

\noindent\textit{Proof.} Using the classical definition of Schur polynomials we obtain
\begin{eqnarray}
s_{\lambda}(1/y) = \frac{\det(y_i^{1-j-\lambda_{N-j+1}})}{\det(y_i^{1-j})}= \frac{\det(y_i^{N-j+\lambda_1+\lambda_N-\lambda_{N-j+1}})}{\det(y_i^{N-j})} \, \prod_{i=1}^N y_i^{-\lambda_1-\lambda_N}. \nonumber
\end{eqnarray}

Since\footnote{As usual, we define $\lambda_i=0$ whenever the index $i$ is greater than $\lambda$'s length.} $\lambda^N_i=\lambda_1+\lambda_N-\lambda_{N-i+1}$, it follows that
\begin{eqnarray}
s_{\lambda}(1/y) = \frac{\det(y_i^{N-j+\lambda^N_j})}{\det(y_i^{N-j})} \, \prod_{i=1}^N y_i^{-\lambda_1-\lambda_N} = \frac{\det(y_i^{j-1+\lambda^N_{N-j+1}})}{\det(y_i^{j-1})} \, \prod_{i=1}^N y_i^{-\lambda_1-\lambda_N}=s_{\lambda^N}(y) \, \prod_{i=1}^N y_i^{-\lambda_1-\lambda_N}, \nonumber
\end{eqnarray}
as we wanted to prove. \qed

Proposition \ref{prop9}, when combined with Lemma \ref{lemma5}, allows us to prove the following.

\begin{coro}
Let $\lambda$ be any partition. Then,
\begin{eqnarray}
\label{iguri9}
&& J'(\lambda^N) = J(\lambda) \, \prod_{i=1}^{N} \frac{\mathbf{s}(a+(N-i)\rho)}{\mathbf{s}(a+b+(2N-i-1)\rho)}, \nonumber
\end{eqnarray}
where the dash means that the parameter $a$ must be changed to $a'=1-a-b-2(N-1)\rho-{\lambda}_1-{\lambda}_N$.
\end{coro}

\noindent\textit{Proof.} Iterating Eq.~(\ref{iguri8}) $N$ times we obtain
\begin{eqnarray}
\label{iguri9bis}
&& J(\lambda) =J_N(\lambda)= J_0(\lambda) \prod_{i=1}^{N} \frac{\mathbf{s}(a+b+(2N-i-1)\rho)}{\mathbf{s}(a+(N-i)\rho)}. \nonumber
\end{eqnarray}
The change of variables $y_i \mapsto 1/y_i$ in $J_0(\lambda)$ and Eq.~(\ref{schur1/y}) prove the corollary. \qed

Finally, let us prove the following.

\begin{teo}
Let $\lambda$ and $\bar{\lambda}$ be two partitions. Then,
\begin{eqnarray}
\label{iguri2}
&& \mathcal{J}(\lambda,\bar{\lambda}) = \int s_{\lambda}(z) \, s_{\bar{\lambda}}(\bar{z}) \, \Phi(z) \Phi(\bar{z}) \, dz \, d\bar z \nonumber \\
&& ~~~~~ = \frac{1}{N!} \, J(\lambda) J(\bar{\lambda}) \prod_{i=1}^{N} \frac{\mathbf{s}(a+(N-i)\rho) \, \mathbf{s}(b+(N-i)\rho) \,  \mathbf{s}(i\rho)}{\mathbf{s}(a+b+(2N-i-1)\rho) \, \mathbf{s}(\rho)}. \nonumber
\end{eqnarray}
\end{teo}

\noindent\textit{Proof.} We will start transforming $\mathcal{J}(\lambda,\bar{\lambda})$ into a multiple contour integral as in \cite{dotsenko2}. Let us introduce $z_i=u_i +i v_i$ for $i=1,2,\dots,N$, $u_i$ and $v_i$ being real variables, and write
\begin{eqnarray}
&& \mathcal{J}(\lambda,\bar{\lambda})= \int s_{\lambda}(u + i v) s_{\bar{\lambda}}(u - i v) \prod_{i=1}^N (u_i^2 + v_i^2)^{a-1} [(1-u_i)^2+v_i^2]^{b-1}  \nonumber \\ 
&& ~~~~~ \times \prod_{i<j}^N [(u_i-u_j)^2+(v_i-v_j)^2]^{2\rho} \, du \, dv. \nonumber
\end{eqnarray}

After analytical continuation (see \cite{aomoto2} for the details), integration contours of the $v$'s can be shifted close to the imaginary axis, i.e., we can make the following change of variables: $v_i\mapsto -i \exp (-2i\epsilon)v_i$, $\epsilon$ being a vanishingly small positive number. We obtain
\begin{eqnarray}
&& \mathcal{J}(\lambda,\bar{\lambda}) \sim \int s_{\lambda}(u + e^{-2i \epsilon}v) \, s_{\bar{\lambda}}(u -e^{-2 i\epsilon} v) 
\prod_{i=1}^N (u_i^2 -  e^{-4 i \epsilon} v_i^2)^{a-1}  \nonumber \\
&& ~~~~~ \times [(1-u_i)^2-e^{-4i\epsilon} v_i^2]^{b-1} \prod_{i<j}^N [(u_i-u_j)^2- e^{-4i\epsilon} (v_i-v_j)^2]^{2\rho} \, du \, dv, \nonumber
\end{eqnarray}
where $\sim$ means that the identity is up to factors that do not depend neither on $\lambda$ nor on $\bar{\lambda}$. Actually, it is up to a phase factor of the form $e^{-2is\epsilon}$ which is irrelevant since the limit $\epsilon \rightarrow 0^+$ will be performed after the integration. It is also up to a factor depending on $N$.

A final change of variables: $y_i=u_i+v_i$, $w_i=u_i-v_i$ allows us to write
\begin{eqnarray}
&& \mathcal{J}(\lambda,\bar{\lambda}) \sim \int s_{\lambda}(y - i \epsilon (y-w)) s_{\bar{\lambda}}(w + i\epsilon (y- w)) \prod_{i=1}^N (y_i - i \epsilon (y_i-w_i))^{a-1} \nonumber \\
&& ~~~~~ \times (w_i + i \epsilon(y_i-w_i))^{a-1} (1-y_i+i\epsilon (y_i-w_i))^{b-1} (1-w_i-i\epsilon (y_i-w_i))^{b-1} \nonumber\\
&& ~~~~~ \times \prod_{i<j}^N (y_i-y_j-i\epsilon(y_i-w_i+y_j-w_j))^{2\rho} (w_i-w_j+i\epsilon(y_i-w_i+y_j-w_j))^{2\rho} \, dy \, dw,
\label{df2}
\end{eqnarray}
which is a double integral that factorizes as a product of two single contour integrals after the limit $\epsilon\rightarrow 0^+$ is performed.

The $\epsilon$-dependence of the integral tells us how the integration contours should be deformed in order to avoid singularities and keep them away from each other: when $y_i < y_j$ the contour of $w_i$ must lie below the one of $w_j$. The integral (\ref{df2}) can therefore be rewritten as
\begin{eqnarray}
&& \mathcal{J}(\lambda,\bar{\lambda}) \sim \sum_{\sigma} I_{\sigma}(\lambda) \, J_{\sigma}(\bar{\lambda}), \nonumber
\end{eqnarray}
where $\sigma$ runs over all possible orderings of the $y$'s contours, $I_{\sigma}(\lambda)$ denotes the integrals over the $y$'s ordered according to $\sigma$, and $J_{\sigma}(\bar{\lambda})$ denotes the contour integrals of the $w$'s following the prescription we have already described.

If one or more of the $y$'s lie outside the interval $(0,1)$ then at least one of the contours of the $w$'s can be deformed to infinity making the integral vanish. On the other hand, since Schur polynomials are symmetric, the integration limits in $I_{\sigma}(\lambda)$ can be freely set to $0$ and $1$. This shows that $I_{\sigma}(\lambda)$ does not depend on $\sigma$ but only on $N$. Actually, we have, up to a factor depending on $N$ and $\rho$,
\begin{eqnarray}
I_{\sigma}(\lambda)\sim J(\lambda)=\int s_{\lambda}(y) \, \Phi(y) \, dy. \nonumber
\end{eqnarray}

On the other hand, we have that $J_{\sigma}(\bar{\lambda})$ is given by
\begin{eqnarray}
J_{\sigma}(\bar{\lambda}) \sim \oint s_{\bar{\lambda}(w)} \, \prod_{i=1}^s w_i^{a-1} (1-w_i)^{b-1} \prod_{i<j} (w_i-w_j)^{2\rho} \, dw,
\label{dff3}
\end{eqnarray}
each integration contour coming from $+\infty$ in the lower half complex plane and going back to $+\infty$ in the upper half complex plane while encircling 
the singularity at $1$ clockwise and not intersecting one with the other. The factor $\prod_{i<j} (w_i-w_j)^{2\rho}$ is defined so that if all the $w_i$ are placed on the real axis and decreasingly ordered, then the phases of the multivalued products all vanish and if the $w_i$ are continued along their contours and are respectively taken around some other point, say $w_j$, in such a way that the contour of $w_i$ goes above $w_j$, the product $\prod_{i<j} (w_i-w_j)^{2\rho}$ gets an additional phase factor $e^{-2\pi i \rho}$. It follows that (\ref{dff3}) can be identified, up to a phase factor depending neither on $\lambda$ nor on $\bar{\lambda}$ (actually, it only depends on $N$ and $\rho$), with 
\begin{eqnarray}
\label{90}
J_{\sigma}(\bar{\lambda}) \sim \int_{1}^{\infty} s_{\bar{\lambda}}(w) \, \Phi(w) \, dw.
\end{eqnarray}

Now, let us perform the change of variables $w_i \mapsto 1/w_i$ in (\ref{90}). Using the inversion identity (\ref{schur1/y}) we conclude that $J_{\sigma}(\bar{\lambda})$ equals (again, up to a phase factor, now depending also on $b$) $J'(\bar{\lambda}^{N})$. 

Summarizing, we have proved that $\mathcal{J}(\lambda,\bar{\lambda})$ is proportional to the product $J(\lambda)J'(\bar{\lambda}^N)$, the proportionality constant being independent of $\lambda$ and $\bar{\lambda}$. The independence of this factor upon the partitions allows us to read it from Dotsenko-Fateev integral putting $\lambda=\bar{\lambda}=0$. Using Lemma \ref{lemma5} we finally prove the theorem. \qed


\begin{coro} Let $\lambda=(2^n1^{m-n})$ and $\bar{\lambda}=(2^{\bar n}1^{\bar m - \bar n})$. Then,
\begin{eqnarray}
&& \mathcal{J}(\lambda,\bar{\lambda}) = \frac{1}{N!} \, m_{\lambda}(1^N) \, m_{\bar{\lambda}}(1^N) \, J(0)^2 \, \frac{[a+(N-1)\rho]^{(\rho)}_{\lambda}}{[a+b+2(N-1)\rho]^{(\rho)}_{\lambda}} \, \frac{[a+b+(N-2)\rho]^{(\rho)}_{n}}{[a+b+(2N-m-2)\rho]^{(\rho)}_{n}} \nonumber \\
&& ~~~~~ \times  \frac{[a+(N-1)\rho]^{(\rho)}_{\bar{\lambda}}}{[a+b+2(N-1)\rho]^{(\rho)}_{\bar{\lambda}}} \, \frac{[a+b+(N-2)\rho]^{(\rho)}_{\bar{n}}}{[a+b+(2N-\bar m-2)\rho]^{(\rho)}_{\bar n}} \, \prod_{i=1}^{N} \frac{\mathbf{s}(a+(N-i)\rho) \, \mathbf{s}(b+(N-i)\rho) \,  \mathbf{s}(i\rho)}{\mathbf{s}(a+b+(2N-i-1)\rho) \, \mathbf{s}(\rho)} \nonumber \\
&& ~~~~~ \times {_4}F_3\left[ \begin{array}{c}
-n, \, -N+m, \, \alpha+\beta+\gamma+2N-n-1, \, \alpha+N-n+1 \\ \alpha+\beta+N-n-1, \, \alpha+\gamma+N-n, \, m-n+2 \end{array} \right] \nonumber \\
&& ~~~~~ \times {_4}F_3\left[ \begin{array}{c}
-\bar n, \, -N+\bar m, \, \alpha+\beta+\gamma+2N-\bar n-1, \, \alpha+N-\bar n+1 \\ \alpha+\beta+N-\bar n-1, \, \alpha+\gamma+N-\bar n, \, \bar m-\bar n+2 \end{array} \right]. \nonumber
\end{eqnarray}
\end{coro}

\end{document}